\documentstyle[12pt]{article}
\def\x{{\mbox{\boldmath$x$}}}
\def\u{{\mbox{\boldmath$u$}}}
\def\r{{\mbox{\boldmath$r$}}}

\def\p{{\mbox{\boldmath$p$}}}

\def\eps{{\epsilon}}
\def\pt{{\tilde p}}
\def\pb{{\bar {p}}}

\def\begineq{\begin{equation}}
\def\endeq{\end{equation}}

\begin{document}
\bibliographystyle{prsty}
\title{Bottleneck effects in turbulence: Scaling phenomena in r- versus
p-space}
\author{Detlef Lohse $^{1*}$ and Axel M{\"u}ller-Groeling $^{2\dagger}$}
\maketitle

\centerline{$^1$ The James Franck Institute, The University of Chicago,}
\centerline{ 5640 South Ellis Avenue, Chicago, IL 60637, USA}

\vspace{.5cm}

\centerline{$^2$ Department of Physics, University of Toronto,}
\centerline{ 60 St.George Street, Toronto, Ontario M5S 1A7, Canada}

\bigskip

\date{}
\maketitle
\bigskip
\bigskip
\bigskip

We (analytically) calculate the energy spectrum corresponding to various
experimental and numerical turbulence data analyzed
by Benzi et al. \cite{ben94a}.
We find two bottleneck
phenomena:
While the local scaling exponent $\zeta_r(r)$ of the structure function
decreases monotonically, the local scaling
exponent $\zeta_p(p)$ of the corresponding spectrum has a minimum
of $\zeta_p(p_{min})\approx 0.45$ at
$p_{min}\approx (10 \eta)^{-1}$ and a maximum
of $\zeta_p(p_{max})\approx 0.77$ at
$p_{max}\approx 8 L^{-1}$.
A physical argument starting from the constant energy flux in p--space
reveals the general mechanism underlying the energy pileups at both ends
of the p--space scaling range.
In the case studied here, they are induced by viscous dissipation and the
reduced spectral strength on the scale of the system size, respectively.

\vspace{0.5cm}\noindent
PACS: 47.27.-i


\newpage
In nonlinear dynamics
scaling exponents in r-space and p-space are often
identified   with each other.
In the case of fully developed turbulence
the scaling exponent $\zeta_{r,2}$ of the velocity structure function
$D^{(2)} (r) = \langle (\u(\x+\r) - \u(\x))^2\rangle \propto r^{\zeta_{r,2}}$
is believed to coincide with the scaling exponent $\zeta_{p,2}$ of the energy
spectrum
of the velocity field (multiplied by $p$), $E(p)\propto
p^{-\zeta_{p,2}-1}$. In this letter we demonstrate that this identification
only holds in the limit of very large Reynolds number. For
(Taylor)-Reynolds numbers $Re_\lambda < 200$
typically achieved in
full numerical simulations \cite{she93}
bottleneck phenomena \cite{fal94} lead to considerable differences between
$\zeta_{r,2}$ and $\zeta_{p,2}$ .
First, for large $p$ near the (inverse) scale of dissipation,
the bottleneck effect accounts
for the puzzling observation that the
numerical spectrum is clearly {\it flatter}
both in experiment \cite{she93b,zoc94} and in numerics
\cite{she93,por94}
than $E(p) \propto
p^{-5/3}$ instead of being steeper as expected from the possibility
of intermittency corrections.
Second, for small $p$ near the inverse external length scale $L^{-1}$
(where $L$ is defined by the driving force), a
similar bottleneck effect leads to {\it steeper} spectra.
This is another hint that {\it finite size effects} as also found and
analyzed in \cite{gnlo94a,gnlo94b,sir94,yak94}
have to be considered.
Our observation has far--reaching consequences for both the numerical and
experimental determination of asymptotic scaling exponents
from spectra.

We first focus on the crossover between viscous subrange (VSR) and
inertial subrange (ISR) and
start from
Benzi et al.'s \cite{ben94a}
{\it measured } longitudinal \cite{my75}
velocity structure
function, assuming that the system  size $L \to \infty$
\cite{ben94a}, i.e., no large scale
finite size effects \cite{gnlo94a,amg94a} are considered.
Benzi et al.\ \cite{ben94a}
analyzed various numerical and experimental data by means of
the extended self similarity method \cite{ben93b,ben94a} and
found that for
$r>\eta$ ($\eta$ being the Kolmogorov scale) the $m^{th}$ longitudinal velocity
structure function $D_L^{(m)}(r)$ obeys
$
D_L^{(m)} (r) = C_m \left( r f(r/\eta)\right)^{\zeta_{r,m}} ,
$
with a {\it universal} function $f(r/\eta)$ for all moments $m$, for
all $Re_\lambda$, and for all kinds of isotropic flow.
We
restrict ourselves to the second order structure functions
and drop the index 2 in what follows.

The structure functions $D_L(r)$ and $D(r)$ are connected by \cite{my75}
$
D (r) = 3 D_L (r) + D_L(r) d\ln D_L(r) / d\ln r.
$
Both functions can be fitted by parametrizations of the Batchelor type
\cite{bat51,my75}.
Originally given by Batchelor as a parametrization,
this formula (1) recently got
theoretical support
by Sirovich, Smith, and Yakhot
\cite{sir94}, who moreover find agreement between
the Batchelor energy spectrum
 and numerical
spectra \cite{chen93,she93} for 30 orders of magnitude.
The high quality of the Batchelor
fit has also been established by older experiments,
for an overview see \cite{my75,eff87}.
Here, the Batchelor fit for $D(r)$,
\begineq
{D(r)\over v_\eta^2} = { r^2 / (3\eta^2)
 \over \left( 1 + \left({1\over 3b}
\right)^{3/2}  \left( {r\over \eta } \right)^2 \right)^{1-\zeta / 2}},
\label{eq4}
\endeq
is slightly superior to that of $D_L(r)$, see fig.\ 1.
Here, $\zeta$ denotes
the asymptotic value of $\zeta_r$ for $r\gg \eta$,
and $v_\eta$ and $b$ are the
Kolmogorov velocity and
Kolmogorov constant \cite{my75}, respectively.
The
experimental value $f(r=\eta) = 8.577 \cdot 10^{-3} = f(1)$ determines
$b=11/(45 (f(1))^{2/3})=5.834$, slightly smaller than $b\approx 6.0 - 8.4$
found in
older experiments, which also show excellent
agreement with (\ref{eq4}) \cite{my75,eff87}.
We determine $D(r)$ from a spline fit to the data and compare the result
with the Batchelor parametrization (\ref{eq4}) in fig.1.
There are no visible deviations.

The {\it local} logarithmic slope \cite{gnlo94a} of eq.\ (\ref{eq4}),
\begineq
\zeta(r) = {d\ln D(r)\over d\ln r} = 2-{(2-\zeta)r^2\over r_d^2 + r^2 }
\label{eq5}
\endeq
is {\it monotonically decreasing} for increasing $r$.
Here, $r_d=(3b)^{3/4}\eta $ (for $\zeta=2/3$)
determines the r--space crossover, defined by equating
the limits for large and small $r$
of eq.\ (\ref{eq4}),
$(r_d/\eta )^2/3 = b(r_d/\eta)^{2/3}$.

Next, we calculate the spectrum $E(p)$ which is, when
neglecting boundary terms,  given by
\cite{my75,amg94a}
\begineq
E(p) = - {1\over 2\pi} \int_0^\infty pr { \sin(pr)}
 D(r) dr .
\label{eq6}
\endeq
In view of our results in fig.1 we feel justified to consider eq.\ (\ref{eq4})
as an exact description of the
experimental structure function of ref.\cite{ben94a}.
Inserting eq.\ (\ref{eq4}) into eq.\ (\ref{eq6}) we obtain
\begin{eqnarray}
E(p) &=& - {pr_d v_\eta^2\over 12\pi\eta^2} {d^3\over
dp^3}\int_{-\infty}^{\infty}
{\exp{(ipr_dx)} \over (1+x^2)^{1-\zeta/2}} dx \nonumber\\
     &=& {r_d^3 2^{-1+\zeta/2} v_\eta^2\over 3
\sqrt{2\pi}\Gamma(1-\zeta/2)\eta^2}
\left(
\zeta\pt^{1/2-\zeta/2}K_{3/2+\zeta/2}(\pt)+\pt^{3/2-\zeta/2}K_{1/2+\zeta/2}(\pt)
\right).
\label{eq8}
\end{eqnarray}
Here, $\pt=pr_d=p/p_d$ and
$K_\nu(\pt)$ is the modified Bessel function \cite{abr70}.
A similar Fourier transformation
of the longitudinal structure function was performed by Sirovich, Smith,
and Yakhot \cite{sir94}.
\footnote{
When the transcriptional error in
eq.\ (20) of ref.\ \cite{sir94}
is corrected,
the bottleneck pileup also shows up. }.
Expanding eq.\ (\ref{eq8}) for small $\pt<1$ and $\zeta>0$ gives
\begineq
E(\pt) = {r_d^3 2^{\zeta-1/2} \zeta v_\eta^2 \over 3 \sqrt{2\pi}\eta^2}
         {\Gamma(3/2+\zeta/2) \over \Gamma(1-\zeta/2)}
\pt^{-\zeta-1}\left(1+ {2-\zeta\over 2\zeta (1+\zeta
  )}\pt^2 + ... \right),
\label{eq9}
\endeq
i.e., we have a {\it positive} correction term to the expected asymptotic
scaling
$E(p)\propto p^{-\zeta -1}$.
This correction signals the onset of
an {\it energy pileup} around $p_d$, see
fig.\ 2.
For large $\pt\gg 1$ the spectrum decays as $E(\pt) \propto
\pt^{1-\zeta/2} \exp{(-\pt)}$.
Fig.2 also shows a frequently used parametrization \cite{foi90} for $E(p)$,
\begineq
E(p) = c p^{-\zeta -1 }\exp{(-p/p_d')},
\label{eq11}
\endeq
where $p_d'$ is chosen in such a way that the r-space crossover $r_d'$
corresponding to eq.\ (\ref{eq11}) coincides with $r_d$, for details see
ref.\cite{amg94a}. This comparison emphasizes the energy pileup around $p_d$
described by the (modified) Sirovich-Smith-Yakhot formula
eq.\ (\ref{eq8}), which can be
considered to be an {\it experimental} spectrum summarizing the various
simulations and experiments of ref.\ \cite{ben94a} and also those
summarized in \cite{eff87}.

As already stated above, the energy pileup leads to
a {\it non
monotonous} local slope
\begin{eqnarray}
\lefteqn{
{d\ln E(p) \over d\ln p} = -\zeta_p(p) -1 } \nonumber\\
& &=
{[-\zeta(1+\zeta)\pt^{1/2-\zeta/2}-\pt^{5/2-\zeta/2}]K_{3/2+\zeta/2}(\pt)
+(2-\zeta)\pt^{3/2-\zeta/2}K_{1/2+\zeta/2}(\pt )\over
\zeta\pt^{1/2-\zeta/2}K_{3/2+\zeta/2}(\pt)+\pt^{3/2-\zeta/2}
K_{1/2+\zeta/2}(\pt)}.
\label{eq10}
\end{eqnarray}
For $\zeta=2/3$ the maximum
local slope is $-1.448$ (instead of $-5/3$) and occurs at
$p_{min} \approx 0.85 p_d  \approx (10\eta)^{-1}$.
Fig.\ 3 shows $\zeta_p(p)$ together with
$\zeta_r(r=1/p)$ from eq.\ (\ref{eq5}),
demonstrating the strikingly different behavior of the local slopes in
r-- and in p--space.

The energy pileup around $p_d$ has also been observed in further
experiments \cite{she93b}
(fitted by a correction term $\propto
p^{2/3}$ instead of our $p^2$, cf. eq.\ (\ref{eq9})),
in full numerical simulations
\cite{she93,por94}, and in a reduced wave vector set approximation
(REWA) of the
Navier-Stokes equations \cite{gnlo94b}.
In ref.\cite{gnlo94b} a correction term
$+2(p/p_{peak})^{1.8}$ was fitted to the data in nice agreement with
our present result, $+3\pt^2/5 = + 2.6 (p/p_{peak})^2$, where $p_{peak}$
is the point of maximum energy dissipation.
Falkovich \cite{fal94} introduced the name ``bottleneck phenomenon''
for the energy pileup
and predicts the correction term to be $\propto (p/p_d)^{4/3}/
\log(p_d/p)$.

We offer the following {\it physical}
explanation (already given in \cite{gnlo94b}) of the bottleneck phenomenon:
Consider the turbulent energy transfer
downscale,
$T(p) \sim p u(\p)\int dp_1 dp_2 u(\p_1)u(\p_2)
\delta(\p + \p_1+\p_2)$, which does not depend on $p$ in the inertial
range due to Kolmogorov's structure equation \cite{my75}. Assume
that the amplitudes $u(\p_1)$, $u(\p_2)$ with $p_1$, $p_2>p_d >  p$
are already damped by viscosity. Then the energy transfer $T(p)$ would be
reduced and stationarity could not be achieved unless $u(\p)$
increases. Because of the locality of the Navier-Stokes interaction in
p-space, the effect is strongest around $p_d$, leading to the energy
pileup. Of course there is also viscous damping, but for
$p <  p_d<\eta^{-1}$ it is smaller than the eddy viscosity $T(p)$
\cite{fal94}. Borue and Orszag's simulations \cite{por94} indeed show
that the pileup starts in a region where $T(p)$ is still constant.
Above explanation rules out spectra of type
(\ref{eq11}). For an explanation of the bottleneck effect within the test
field model we refer to ref.\ \cite{her82}, see also \cite{qia84}.
For an analogous phenomenon in temperature spectra see \cite{tat92}.

{\it Formally} the bottleneck phenomenon reflects the relatively sharp
crossover from $r^2$-scaling (VSR) to $r^\zeta$-scaling (ISR)
in the structure function (\ref{eq4}).
To illustrate this we transform the spectrum  (\ref{eq11}) back to r--space.
This spectrum does not show the bottleneck
phenomenon and the corresponding structure function
\begineq
D(r)=  {4 c\Gamma(-\zeta ) \over r(\zeta +1 )
p_d^{\prime\zeta + 1}} \left(p_d'r (\zeta + 1) -\left(1+p_d^{\prime2}r^2
\right)^{(\zeta + 1)/2} \sin{\left( (\zeta +1) \arctan (p_d'r)
\right)}\right)
\label{eq12}
\endeq
differs from the Batchelor parametrization (\ref{eq4}) by its considerably
{\it smoother} transition (see the dotted curve in fig.1,
showing a ratio of  $\approx$ 1.8  around $r_d$.).

Our explanation suggests that the bottleneck effect potentially accompanies
any
sudden change in
spectral strength, provided the wave vector amplitudes
interact nonlinearly and a conserved flux exists. We are consequently
led to expect a similar effect at the infrared
end of the scaling regime where the small--$p$ modes are reduced
in their spectral strength by the
finite system size.

Let us therefore consider the crossover between ISR and the large $r$
saturation domain, where $D(r)=2\langle \u^2 \rangle = 6 u^2_{1,rms}$
becomes constant. Recall that $L\equiv 1/p_L$ is the forcing scale.
{} From experimental data \cite{my75,ben94a,eff87} we
conclude that the second crossover at $r = L$ is again
well described
by a Batchelor type transition \footnote{
This crossover is probably nonuniversal. The important point here is
simply the reduced spectral strength for small $p$,
induced by the finite size, i.e.,
$E(p) \rightarrow 0$ as $p\rightarrow 0$.},
\begineq
D(r) = 2 \langle \u^2 \rangle r^2 \cdot (r_d^2 + r^2)^{-1+\zeta/2}
\cdot (L^2 + r^2 )^{-\zeta/2},
\label{eq15}
\endeq
see fig.\ 1.
The general mechanism outlined above should equally well apply in this regime:
The velocity
amplitudes of the modes $p_1$, $p_2<p_L < p$ (or either of them) are
reduced because of the finite size of the system. The mode $u(\p)$
again has to increase in order to guarantee a p-independent energy
flux,
now resulting in a {\it steeper} spectrum.

Indeed, we find such a behavior for the spectrum corresponding to (\ref{eq15}).
For $r_d \ll r$ we derive the
analytical result (for $\zeta =2/3$)
\begin{eqnarray}
E(p)  &=&  {\langle \u^2 \rangle L \over \pi }
\Bigg( - {\Gamma (5/6) \over \Gamma(1/3)} \sqrt{\pi}
\left[
{5\over 9} \pb^2
\phantom{}_1F_2({11\over 6},{5\over 2}, {5\over
  2},{\pb^2\over4})
+
{11\over 405} \pb^4
\phantom{}_1F_2({17\over 6},{7\over 2}, {7\over
  2},{\pb^2\over4}) \right] \nonumber \\
&+& {\pi \over 2}
\left[
{1\over 3} \pb
\phantom{}_1F_2({4\over 3},{2}, {3\over
  2},{\pb^2\over4})
+
{2\over 27} \pb^3
\phantom{}_1F_2({7\over 3},{3}, {5\over
  2},{\pb^2\over4}) \right]
\Bigg)
\label{eq16}
\end{eqnarray}
where $\pb = p/p_L$ and $\phantom{}_1F_2(a,b,c,z)$
denotes a generalized hypergeometric function
\cite{abr70}.
The spectrum and the corresponding $\zeta_p (p)$
are shown in the left parts of figs.\ 2 and 3, respectively.
We find $p_{max} \approx 8 p_L$ and $\zeta (p_{max}) \approx
0.77$. Thus the deviations from classical scaling are again much
larger than the discussed intermittency corrections. Note that our
result agrees with theoretical \cite{gnlo94a,yak94} and
experimental  hints (summarized in \cite{yak94})
that the spectra are steeper for
small $p$.

We finally calculate the {\it effective} scaling exponent
$\zeta_p^{(eff)} (Re_\lambda)$ that will be measured in p-space
simulations. Here we only consider
the bottleneck phenomenon for large $p$, as in most numerical schemes
the smallest wave vectors are forced and no $p < p_L$ are included.
Let us express $r_d$
in terms of $L$ and the  Taylor-Reynolds number
$Re_\lambda = \lambda
u_{1,rms}/\nu$, where $\lambda = u_{1,rms } / (\partial_1 u_1)_{rms}$
is the Taylor length.
We have $\eps = c_{\eps}
u^3_{1,rms}/L$ with $c_\eps = (6/b)^{3/2} \approx 1$,
which is also known from grid turbulence
experiments \cite{sre84}.
On the other hand,
$\eps = 15\nu (\partial_1u_1)^2_{rms}$ \cite{my75}.
Using these relations we finally get
$\eta = 15^{3/4} c_\eps^{-1} L Re_\lambda^{-3/2}$  or (for $\zeta=2/3$)
$r_d = (3b)^{3/4} \eta \approx  63 L Re_\lambda^{-3/2}.$
This connection between $r_d/L$ and $Re_\lambda$ allows us to calculate
$\zeta_p^{(eff)} (Re_\lambda )$ as the average
\begineq
\zeta_p^{(eff)} (Re_\lambda )
= {1\over \ln(p_{min}/p_L)} \int_{p_L}^{p_{min}} \zeta_p(p) d\ln p,
\label{eq14}
\endeq
where $p_{min}$ is, as above, the wave vector of minimal $\zeta_p(p)$.
The function
$\zeta_p^{(eff)} (Re_\lambda )$
is shown in the inset of fig.\ 3. The deviations from the asymptotic value
$\zeta_p^{(eff)} (Re_\lambda\to \infty )=\zeta$ are large. Assuming
$\zeta=2/3$, even for the largest $Re_\lambda =200$ achieved in
numerical simulations \cite{she93} we have
$\zeta_p^{(eff)} \approx 0.58$,
which very well agrees with what is observed in numerical simulations
\cite{she93}. Impressive {\it experimental} confirmation
of our prediction follows from recent measurements by Zocchi et al.
\cite{zoc94}. We include their
data for $\zeta_p^{(eff)} (Re_\lambda)$ in our figure.


Let us finally remark that
our physical explanation of the bottleneck energy pileups is very
general, it only assumes some inertial range with a constant energy
flux  in p-space. E.g., these conditions hold for surface or
capillary waves \cite{zak92}, where bottleneck phenomena are also
expected \cite{fal94}, or for KS dynamics \cite{kur78}.
How  bottleneck phenomena manifest themselves
in higher order moments and in power spectra
remains a question for further research.

\vspace{1.5cm}
\noindent
{\bf Acknowledgements:}
We are grateful to R.\ Benzi, who kindly supplied us with his
experimental data. We thank him and M. Brenner,
S. Esipov, A. Esser, G. Falkovich, A. Golubentsev,
S. Grossmann, J. Herring, M. Jensen, L. Kadanoff, R. Kerr, S. Kida,
A.\ Praskovsky, and L. Sirovich
for discussions and hints.
D. L.
acknowledges support by a NATO grant
through the Deutsche Akademische Austauschdienst (DAAD),
and by DOE. A.M.--G. was supported by NSERC.

\newpage

\centerline{\bf Figures}
\begin{figure}[htb]
\caption[]{
Velocity structure function $D(r)$,
calculated from Benzi et al.'s data (dashed) \cite{ben94a},
and its Batchelor fit (\ref{eq4}) (solid) for $\zeta=2/3$. Both
curves are identical, which can even be seen in the enlargement of the
crossover.
The dotted line shows a structure function corresponding to
the spectrum  (\ref{eq11}).
The upper right part shows the saturation of the structure function
(\ref{eq15}), dashed.
In the inset the {\it longitudinal} structure function $D_L (r)$ is
shown.
The original data \cite{ben94a}
are diamonds, the dashed line is
a spline fit of these data, the solid line a  fit of Batchelor type.
Slight differences are seen.
}
\label{fig1}
\end{figure}

\begin{figure}[htb]
\caption[]{
Experimental energy spectrum  eq.\ (\ref{eq8}) (solid)
with $\zeta = 2/3$
and eq.\ (\ref{eq11})
(dashed) without the energy pileup.
In the left part the spectrum according to (\ref{eq16}) is shown.
In the insets, the spectrum is enlarged around the energy
pileups and compared
to classical $-5/3$-scaling.
}
\label{fig2}
\end{figure}

\begin{figure}[htb]
\caption[]{
The local p-space scaling exponents $\zeta_p (\pt)$
from  eq.\
(\ref{eq10}) and eq.\ (\ref{eq16}) (solid),
and the local r-space scaling exponent $\zeta_r(rp_d=1/\pt)$ from eq.\
(\ref{eq5}) and (\ref{eq15}) (dashed).
The inset shows
the averaged p-space scaling exponent $\zeta_p^{(eff)} (Re_\lambda )$,
see eq.\ (\ref{eq14}) (solid). Also shown is the local
p-space scaling exponent $\zeta_p(p_L(Re_\lambda ))$, dashed.
We chose $\zeta = 2/3$ throughout.
The dots are the experimental \cite{zoc94} $\zeta_p^{(eff)} (Re_\lambda )$.
In \cite{zoc94} only $\zeta_p^{(eff)} (Re )$ is given, so we calculated
$Re_\lambda = c Re^{1/2}$ with $c=0.17$ chosen to give agreement for
small $Re_\lambda$.
}
\label{fig3}
\end{figure}

\newpage
\noindent
$^*$ On leave of absence from Fachbereich Physik, Universit\"at
Marburg, Renthof 6, D-35032 Marburg.

\noindent
$^\dagger$ Present address:
C.E.A., Service de Physique de l'\'Etat
Condens\'e, Centre d'\'Etudes de Saclay, 91191 Gif sur Yvette Cedex, France.



\begin{thebibliography}{10}


\bibitem{ben94a}
R. Benzi, S. Ciliberto, C. Baudet, and G.~R. Chavarria, ``On the
scaling of 3D, homogenous, isotropic turbulence'', Physica D, in press (1994).


\bibitem{she93}
Z.~S. She, S. Chen,  G. Doolen,  R. H. Kraichnan,  and S. A. Orszag,
Phys. Rev. Lett. {\bf 70},  3251  (1993);
A. Vincent and M. Meneguzzi, J. Fluid Mech. {\bf 225},  1  (1991);
R. Kerr, J. Fluid Mech. {\bf 211},  309  (1990).

\bibitem{fal94}
G. Falkovich, Phys. Fluids {\bf 6},  1411  (1994).

\bibitem{she93b}
Z.~S. She and E. Jackson, Phys. Fluids A {\bf 5},  1526  (1993);
S. G. Saddoughi and S. V. Veeravalli, J. Fluid Mech. {\bf 268}, 333 (1994);
also
C. Wark, 1994, private communication.


\bibitem{zoc94}
G. Zocchi, P. Tabeling, J. Maurer, and H. Willaime, Phys. Rev. E  (1994).

\bibitem{por94}
D. Porter, P. Woodward, and A. Pouquet,
``Inertial range structures in compressible turbulent flows''
preprint, 1994;
V. Borue and S.~A. Orszag,
``Forced 3D Homogenous turbulence with hyperviscosity'',
preprint, 1994.

\bibitem{gnlo94a}
S. Grossmann and D. Lohse, Phys. Fluids {\bf 6},  611  (1994);
S. Grossmann, D. Lohse, V. L'vov, and I. Procaccia, Phys. Rev. Lett. {\bf 73},
  432  (1994);
V.~S. L'vov and I. Procaccia, Phys. Rev. E {\bf 49},  4044  (1994).

\bibitem{gnlo94b}
S. Grossmann and D. Lohse,
``Universality in Turbulence'', Phys. Rev. E  (Oct.\ 1994).

\bibitem{sir94}
L. Sirovich, L. Smith, and V. Yakhot, Phys. Rev. Lett. {\bf 72},  344  (1994).


\bibitem{yak94}
V. Yakhot, Phys. Rev. E {\bf 49},  2887  (1994).




\bibitem{my75}
A.~S. Monin and A.~M. Yaglom, {\em Statistical Fluid Mechanics} (The MIT Press,
  Cambridge, Massachusetts, 1975).

\bibitem{amg94a}
D. Lohse and A. M\"uller-Groeling,
``Anisotropy and scaling corrections in  turbulence'',
Chicago and Toronto, 1994.


\bibitem{ben93b}
R. Benzi, S. Ciliberto, R. Tripiccione, C. Baudet, F. Massaioli, and S. Succi,
Phys. Rev. A {\bf 48},  R29  (1993).

\bibitem{eff87}
H. Effinger and S. Grossmann, Z. Phys. B {\bf 66},  289  (1987).

\bibitem{bat51}
Proc. Camb. Philos. Soc. 47, 359 (1951).



\bibitem{chen93}
S. Chen,  G. D. Doolen,  J. R. Herring,
        R. H. Kraichnan, S. A. Orszag, and Z. S. She,
Phys. Rev. Lett. {\bf 70}, 3051, 1993.


\bibitem{abr70}
M. Abramowitz and I.~A. Stegun, {\em Handbook of Mathematical Functions}
  (Dover, New York, 1970); A.~P. Prudnikov, Yu.~A. Brychkov, O.~I. Marichev,
  {\em  Integrals and Series}, Vol.3 (Gordon and Breach, New York, 1990).




\bibitem{foi90}
C. Foias, O. Manley, and L. Sirovich, Phys. Fluids A {\bf 2},  464  (1990).


\bibitem{her82}
J. R. Herring, D. Schertzer,  M. Lesieur, G. R.
		  Newman, J. P. Chollet, and M. Larcheveque,
J. Fluid Mech. {\bf 124},  411  (1982).

\bibitem{qia84}
J. Qian, Phys. Fluids {\bf 27},  2229  (1984).

\bibitem{tat92}
V.~I. Tatarskii, M.~M. Dubovikov, A.~A. Praskovsky, and M.~Y. Karyakin, J.
  Fluid Mech. {\bf 238},  683  (1992).



\bibitem{sre84}
K.~R. Sreenivasan, Phys. Fluids {\bf 27},  1048  (1984). We neglected the
$Re_\lambda$
dependence of $c_\epsilon$, cf.\ D.\ Lohse, Phys. Rev. Lett. {\bf 73}, Oct.
1994.



\bibitem{zak92}
V.~E. Zakharov, V.~S. L'vov, and G. Falkovich, {\em Kolmogorov Spectra of
  Turbulence} (Springer, Heidelberg, 1992).

\bibitem{kur78}
Y. Kuramoto, Suppl. Prog. Theor. Phys. {\bf 64},  346  (1978);
G. Sivashinsky, Acta Austronautica {\bf 4},  1177  (1977).

\end{thebibliography}
\end{document}